\documentclass[10pt]{article}

\usepackage{amsmath}
\usepackage{amssymb}

\usepackage{graphicx}

\usepackage{cite}

\usepackage{color}

\usepackage[labelfont=bf,labelsep=period,justification=raggedright]{caption}

\makeatletter
\renewcommand{\@biblabel}[1]{\quad#1.}
\makeatother

\date{}

\begin{document}

% Title must be 150 characters or less
\begin{flushleft}
{\Large \textbf{Detecting the Influence of Spreading in Social
Networks with Excitable Sensor Networks} }
% Insert Author names, affiliations and corresponding author email.
\\
Sen Pei$^{1,2}$, Shaoting Tang$^{1,2,*}$, Zhiming Zheng$^{1,2,*}$
\\
$^{1}$ School of Mathematics and Systems Science, Beihang
University, Beijing, China
\\
$^{2}$ Laboratory of Mathematics, Informatics and Behavioral
Semantics, Ministry of Education, Beijing, China
\\
$^{*}$ Corresponding author
\\
E-mail: tangshaoting@buaa.edu.cn (ST), zzheng@pku.edu.cn(ZZ)
\end{flushleft}

% Please keep the abstract between 250 and 300 words
\section*{Abstract}

Detecting spreading outbreaks in social networks with sensors is of
great significance in applications. Inspired by the formation
mechanism of human's physical sensations to external stimuli, we
propose a new method to detect the influence of spreading by
constructing excitable sensor networks. Exploiting the amplifying
effect of excitable sensor networks, our method can better detect
small-scale spreading processes. At the same time, it can also
distinguish large-scale diffusion instances due to the
self-inhibition effect of excitable elements. Through simulations of
diverse spreading dynamics on typical real-world social networks
(facebook, coauthor and email social networks), we find that the
excitable senor networks are capable of detecting and ranking
spreading processes in a much wider range of influence than other
commonly used sensor placement methods, such as random, targeted,
acquaintance and distance strategies. In addition, we validate the
efficacy of our method with diffusion data from a real-world online
social system, Twitter. We find that our method can detect more
spreading topics in practice. Our approach provides a new direction
in spreading detection and should be useful for designing effective
detection methods.

\section*{Introduction}

Plenty of phenomena in various domains can be depicted by the
spreading dynamics in social networks, e.g., the outbreak of a
contagious disease
\cite{Anderson1992,Diekmann2000,Pastor-Satorras2014,Yan2014}, the
diffusion of a piece of information
\cite{Kumar2004,Liben2008,Leskovec2009a,Yan2013}, or the promotion
of a commercial product \cite{Watts2007,Domingos2005,Leskovec2007}.
The detection of spreading processes in social networks is an
important issue in many real-world applications, such as the
formulation of timely intervention measures during the spread of an
epidemic, and the surveillance of current trending topics in popular
online social networks. In recent years, the global outbreaks of
seasonal influenza and the widespread use of online social media in
political movements are reminiscent of the vital role of spreading
detection \cite{Moreno2011}. Because of its practical value,
designing effective detection methods has attracted much attention
across disciplines. In particular, several Internet-based
surveillance systems for disease outbreaks have been proposed
\cite{Ginsberg2008,Culotta2010,Aramaki2011,Wilson2009}. By
monitoring health-seeking behaviors in the form of online search
engine queries or analyzing symptom-related terms appearing in
online social media, these systems can estimate the current level of
spreading activity. Another approach is to deploy social network
sensors in the system. A heuristic algorithm for the optimal
placement of sensors has been proposed for spreading models
\cite{Leskovec2007a}, and it has been shown that properly placed
sensors can detect contagious outbreaks before they happen in large
scale by taking advantage of the informative properties of social
networks \cite{Christakis2010}.  Moreover, recent studies indicate
that the origin of a spreading process can be inferred by placing
sensors in social networks \cite{Lokhov2014,Pinto2012}. In addition,
there exist extensive studies on predicting the evolution of
spreading from a snapshot \cite{Chen2014,Keeling2004,Brockmann2013}.
All these approaches make contributions to the detection and
prediction of spreading processes in social networks.

Most of the existing detection methods focus on the objective of
early detection. However, considering the large numbers of spreading
occurring simultaneously in social networks, it is also important to
detect and distinguish the influence of these spreading processes.
In specific, our goal is to detect as many spreading processes as
possible and rank their relative influence at the same time. This
enables us to have an estimation of these spreading processes, which
can be applied to applications such as selecting and ranking blogs
in a blog community. To achieve this goal, we develop an alternative
approach by making use of excitable sensors. In the field of
psychophysics, it has been well established that the cooperative
effect of excitable elements can be used to explain how physical
stimuli (sound, light and pressure) transduce into psychological
sensations \cite{Kinouchi2006}. Although each single excitable node
responds to stimuli with a small range, the collective response of
the entire excitable system can encode stimuli spanning several
orders of magnitude, yielding high sensitivity and a broad dynamic
range (the range of stimulus intensities resulting in
distinguishable network responses)
\cite{Kinouchi2006,Larremore2011}. Borrowing the property of
nonlinear amplification of stimuli, the excitable elements are
suitable for detecting the influence of spreading in social
networks: the stimuli can be regarded as infections, and the
response of the excitable system can be treated as the detected
influence.

Following this idea, we propose a method to deploy excitable sensor
networks in social systems. For both homogeneous and heterogeneous
networks, we analytically derive the relationship between the
response and spreading influence. Through simulations on
Erd\"{o}s-R\'{e}nyi (ER) and Barab\'{a}si-Albert (BA) networks, we
verify our theoretical analysis for susceptible-infected-recovered
(SIR) model, susceptible-infected-susceptible (SIS) model, rumor
spreading (Rumor) model and susceptible-infected-recovered model
with limited contacting ability (SIRL)
\cite{Hethcote2000,Pei2013,Moreno2004,Yang2007,Yang2008}. Then, we
compare the performance of excitable sensor networks with several
commonly used sensor placement strategies. The simulation results of
spreading models on three typical social networks (including
facebook, coauthor and email social networks) suggest that excitable
sensor networks outperform the other considered methods: excitable
sensor networks can not only grasp small-scale spreading processes
because of the amplifying effect of collective dynamics, but also
better distinguish large-scale diffusion instances due to the
self-inhibition effect of excitable elements. Under the same
circumstances, our method has a larger dynamic range, which means it
can detect spreading in a wider range of influence. In addition, we
discuss the impact of the construction method of excitable sensor
networks. We find that the homogeneous sensor networks perform
better than heterogeneous ones, and the choice of sensors' number
will not affect our results significantly.  We also explore
spreading processes originating from different sources. The
amplifying effect of excitable sensors is found to be more effective
for well-connected sources. Therefore, excitable sensor networks are
more likely to detect spreading instances from high-degree sources.
Furthermore, we validate the efficacy of our method using real
diffusion data from Twitter. We track the spreading of $309$ topics
in Twitter among selected users, and use each method to detect them.
The results show that the excitable sensor network can detect more
spreading topics than other strategies under the same situations.

\section*{Materials and Methods}

\subsection*{Datasets}

In the numerical simulations, we utilize three different typical
social networks as the substrates on which spreading processes
occurs. Here we introduce the properties and statistics of each
adopted social network. Notice that we treat all the networks as
undirected.

{\bf Facebook} The social network of facebook contains all of the
user-to-user links from the Facebook New Orleans networks
\cite{Viswanath2009}. If one user is in the friend list of another
user, then an undirected social link is constructed between them. It
contains 63731 nodes and 1545685 links, resulting in an average
degree of 48.5. This dataset is shared at
http://socialnetworks.mpi-sws.org/data-wosn2009.html.

{\bf Coauthor} Based on the DBLP computer science bibliography which
provides a comprehensive list of research papers in computer
science, a coauthorship network is constructed \cite{Yang2012}. In
the network, two authors are connected if they publish at least one
paper together. There are 317080 nodes and 1049866 links in the
network, and the average degree is 6.7. Researchers can download
this dataset at http://snap.stanford.edu/data/com-DBLP.html.

{\bf Email} In the Enron email communication network
\cite{Leskovec2009}, nodes are email addresses. If an address $i$
sends at least one email to another address $j$, the graph contains
an undirected edge connecting $i$ to $j$. The network contains 36692
nodes and 183831 links, and the average degree is 10.0. The data
source is http://snap.stanford.edu/data/email-Enron.html.

In the validation of excitable sensor networks, we use real
diffusion data from the microblog platform Twitter. The tweets are
sampled between January 23rd and February 8th, 2011 and are shared
by Twitter (http://trec.nist.gov/data/tweets/). In Twitter, specific
topics are usually labelled by a ``hashtag'', i.e., a word or an
unspaced phrase prefixed with the number sign (``\#''). Therefore,
we can use hashtags to track the spreading of a specific topic in
Twitter. During the collection period, there happened to be a mass
protest on January 25th in Egypt, which was an important event of
the "Arab Spring". In this event, Twitter was used by protesters to
organize the protest and recruit members. Many Twitter users
discussed and shared information about this protest using Twitter.
The most used hashtag related to this protest is ``\#Jan25''. To
obtain the information spreading among users participating in this
protest, we filtered 23712 tweets containing ``\#Jan25'' published
by 7014 different users. Then, we checked all the available tweets
that were published by these users and found 309 distinct topics
appearing more than 10 times. To reconstruct the social network of
the selected users, we have extracted the mention network, where two
users are connected if one user has mentioned another user
(``@username'') at least once. Compared with the follower network,
the mention network stands for stronger social relations because
mentions usually contain personal communications. In this way, we
obtain 10547 social links among 7014 users, and the average degree
is 3.0.

\subsection*{Spreading Models}

In this paper, we have applied four spreading models to verify the
effectiveness of excitable sensor networks. These models were
frequently used in previous research works on spreading dynamics. We
introduce the details of each model here.

{\bf SIR model} The SIR model is suitable to describe the spreading
of a disease with immunity. In the SIR model, each individual is in
one of three states: the susceptible (S), infected (I) and recovered
(R). At each time step, infected nodes infect their susceptible
neighbors with probability $\beta$ and then enter the recovered
state with probability $\mu$, where they become immunized and cannot
be infected again. When there are no more infected individuals in
the system, the fraction of recovered person, or equivalently, the
fraction of people who have ever been infected, is denoted by $M$.
Assume the densities of the susceptible, infected, and recovered at
time $t$ are $s(t)$, $i(t)$, and $r(t)$ respectively. In homogeneous
random networks with average degree $\langle k\rangle$, the dynamics
of SIR model satisfies the following set of coupled differential
equations
\begin{equation}\label{sir}
\left\{
\begin{array}{ll}
  \frac{di(t)}{dt}=\beta \langle k\rangle s(t)i(t)-\mu i(t), \\
  \frac{ds(t)}{dt}=-\beta \langle k\rangle s(t)i(t), \\
  \frac{dr(t)}{dt}=\mu i(t).
\end{array}
\right.
\end{equation}

{\bf SIS model} The SIS model, in which only two states, susceptible
(S) and infected (I), are considered, describes spreading processes
that do not confer immunity on recovered individuals. In the
spreading process, infected individuals infect their susceptible
neighbors with probability $\beta$ and return to S state with
probability $\mu$. As time evolves, the fraction of infected persons
$\rho$ will become steady. We run SIS dynamics for $100$ steps and
take the average infected proportion of last $30$ steps as $\rho$.
In ER random networks with average degree $\langle k\rangle$, the
dynamics can be described by the differential equation
\begin{equation}\label{sis}
  \frac{di(t)}{dt}=\beta\langle k\rangle i(t)(1-i(t))-\mu i(t).
\end{equation}

{\bf Rumor model} In Rumor model, each individual can be in three
possible states: the spreader (S), ignorant (I), and stifler (R).
Spreaders represent nodes that are aware of the rumor and are
willing to transmit it. Ignorant people are individuals unaware of
the rumor. Stiflers stand for those that already know the rumor but
are not willing to spread it anymore. In each time step, the
spreaders contact all their neighbors and turn the ignorant ones
into spreaders with probability $\beta$. If the spreaders encounter
spreaders or stiflers, they will turn to stiflers with probability
$\mu$. The influence of the rumor $M$ is defined as the fraction of
stiflers when there are no more spreaders in the system. In
homogeneous networks with average degree $\langle k\rangle$,
dynamics follows the set of coupled differential equations
\begin{equation}\label{rumor}
\left\{
\begin{array}{ll}
   \frac{di(t)}{dt}=-\beta \langle k\rangle i(t)s(t),  \\
   \frac{ds(t)}{dt}=\beta \langle k\rangle i(t)s(t)-\mu\langle k\rangle s(t)[s(t)+r(t)], \\
   \frac{dr(t)}{dt}=\mu\langle k\rangle s(t)[s(t)+r(t)].
\end{array}
\right.
\end{equation}

{\bf SIRL model} The SIRL model is a modified SIR model, in which
each node is assigned with an identical capability of active
contacts, $L$. It stands for the type of spreading with limited
contacting ability. Compared with the standard SIR model, at each
time step in SIRL model, each infected individual will generate $L$
contacts. Multiple contacts to one neighbor are allowed, and
contacts not between susceptible and infected ones are also counted
just like the standard SIR model. The parameters $\beta$ and $\mu$
represent the infection rate and recover rate respectively. In
homogeneous random networks with average degree $\langle k\rangle$,
the dynamics can be calculated by the set of differential equations
\begin{equation}\label{sirl}
\left\{
\begin{array}{ll}
   \frac{di(t)}{dt}=\beta L s(t)i(t)-\mu i(t), \\
   \frac{ds(t)}{dt}=-\beta L s(t)i(t), \\
   \frac{dr(t)}{dt}=\mu i(t).
\end{array}
\right.
\end{equation}

% Results and Discussion can be combined.
\section*{Results and Discussion}

\subsection*{Construction of the excitable sensor network}

\begin{figure}%[!ht]
\begin{center}
\includegraphics[width=4in]{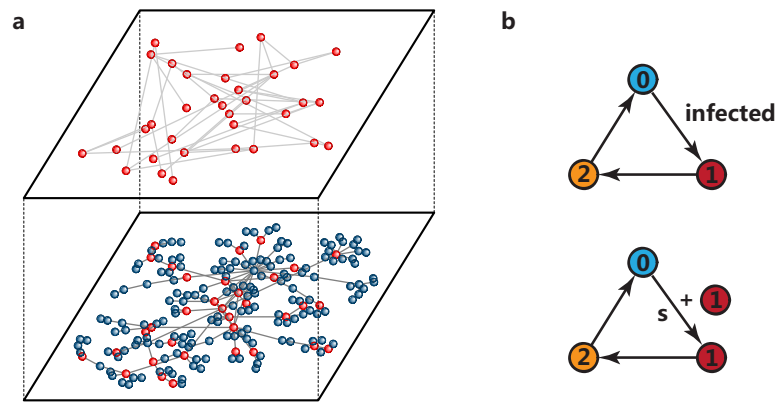}
\end{center}
\caption{ {\bf Illustrations of excitable sensor networks and the
dynamics of sensors.}  (a) A schema of a sensor network. The lower
layer is the underlying social network and the upper layer
represents the sensor network. Both blue and red balls are
individuals in the social network and red balls are selected to be
sensors. The spreading dynamics and signal transmission occur in the
lower and upper level separately. (b) The dynamics of excitable
sensors. The number 0, 1 and 2 stand for the resting, excitation and
refractory state, respectively. Each sensor in resting state can be
activated either by means of infection in the spreading dynamics or
by excited neighboring sensors with probability $s$ independently.
Once activated, the sensors will automatically turn into the
refractory state in the next time step, where they cannot be
activated again and activate other sensors. Then, these sensors
change back to resting state. } \label{fig1}
\end{figure}

We now describe how to construct an excitable sensor network in a
social network. Each excitable node has $n$ states: $s_i=0$ is the
resting state, $s_i=1$ corresponds to excitation, and the remaining
$s_i=2,\ldots,n-1$ are refractory states. In our study, we set
$n=3$. Given a social network, we select $f$ percent of nodes as
sensors according to specific criteria. Then, we create links
between the selected sensors to form a network. In our case, we
construct a homogeneous random network among selected sensors.
Assuming there are $N_s$ nodes in the sensor network, we assign
$N_s\langle k\rangle/2$ links to randomly chosen pairs of nodes,
which produces an average degree $\langle k\rangle$.
Fig.\ref{fig1}(a) illustrates an instance of a sensor network. The
lower layer is the underlying social network, while the upper layer
is the sensor network.

There are two dynamical processes in the system: spreading dynamics
in the social network and signal transmission dynamics in the sensor
network. In our study, we adopt four spreading models that are
frequently used in the research of spreading dynamics: SIR, SIS,
Rumor and SIRL models
\cite{Hethcote2000,Pei2013,Moreno2004,Yang2007,Yang2008}. The SIR
model is developed to describe the contagion process of diseases
with immunity. Once people recover from the disease, they will
acquire permanent immunity. The SIS model depicts the outbreaks of
contagions that an individual can catch more than once. The Rumor
model considers the diffusion process of rumors among a population.
The SIRL model is a modified SIR model, in which each node has a
limited capability of active contacts $L$ at each time step. The
individual-level spreading mechanisms in these models confine the
diffusion processes within the underlying social networks, which
makes them suitable for exploring the interplay between the
spreading dynamics and social network structures. Since the early
days of the research of complex networks, these models have been
widely employed to simulate the epidemic spreading of different
types of diseases
\cite{Pastor-Satorras2014,Barrat2008,Pastor2001,Boguna2002,Yang2007,Yang2008},
information diffusion among individuals
\cite{Kitsak2010,Kleinberg2007,Pei2013,Lu2011}, and rumor
propagation in online social networks \cite{Moreno2004,Borge2012}.
The details of these models are explained in the Materials and
Methods section.

In the following analysis, if an individual catches a contagious
disease or becomes aware of a piece of rumor, we define its state as
infected. For SIR, Rumor and SIRL models, the influence of a
spreading instance $M$ is defined as the proportion of people who
have ever been infected. In contrast, in SIS model, because people
can be infected repeatedly, we define the influence $\rho$ as the
fraction of infected persons when the system reaches a steady state.

Signal transmission dynamics occur in sensor networks. The evolution
dynamics of sensors are shown in Fig.\ref{fig1}(b). At each time
step, sensors in resting state can evolve into active state under
two circumstances: infected in spreading dynamics or activated by
active neighboring sensors with probability $s$ (coupling strength).
The active sensors will update into refractory state and then change
back into resting state deterministically. In refractory state,
sensors can neither be activated nor activate other neighbors. To
define the response of a sensor network to the spreading process, we
assume that the observation time is $T$. Denote the activity level
of sensors $F^t$ at time $t$ as the proportion of active sensors.
The response $F$ is defined as the average activity level during the
observation time, $F=\sum_{t=0}^{T}F^t/T$.

For a spreading process with influence $M$, the system will feedback
a response $F$. As a function of the influence $M$, response has a
minimum value $F_0$ and a maximum value $F_{max}$. To quantify the
ability of excitable sensor networks to detect spreading, we define
the dynamic range $\Delta=10\log_{10}(M_{high}/M_{low})$ as the
range of influence that is distinguishable based on response $F$,
discarding diffusion processes that are too small to be
distinguished from $F_0$ or that are too close to saturation
\cite{Kinouchi2006}. The range $[M_{low},M_{high}]$ is found from
its corresponding response interval $[F_{low},F_{high}]$, where
$F_x=F_0+x(F_{max}-F_0)$. Fig.\ref{fig3}(b) exemplifies the dynamic
range for the interval $[F_{0.1},F_{0.9}]$, where $low=0.1$ and
$high=0.9$.

The dynamics of excitable sensor networks are highly related to the
network topology and coupling strength $s$. Many previous studies
have shown that there exists a critical point in excitable networks.
Only above the critical point, does self-sustained activity emerge
in response to external stimuli. It has been proved that the
critical state occurs when the largest eigenvalue of the interacting
adjacency matrix is exactly $1$ \cite{Larremore2011}. Particularly,
a network of excitable elements has its sensitivity and dynamic
range maximized at the critical point
\cite{Larremore2011,Larremore2011a,Pei2012,Larremore2014}.
Therefore, we select the coupling strength $s$ to make the sensor
network achieve a critical state. For random networks with average
degree $\langle k\rangle$ and coupling strength $s$, the largest
eigenvalue of the interacting adjacency matrix is approximated by
$s\langle k\rangle$. Thus, in designing the excitable sensor
networks, we set the coupling strength $s=1/\langle k\rangle$ to
optimize the dynamic range.

\subsection*{Dynamics of excitable sensors }

The evolution dynamics of excitable sensors can be studied through
theoretical analysis. For each sensor $i=1,\cdots,N_s$, we denote
the probability in active state at time $t$ as $p_i^t$. Then, the
activity level of sensors is $F^t=\sum_{i=1}^{N_s}p_i^t/N_s$. In the
case of small-scale spreading, we assume that sensors are activated
independently by their neighboring active sensors. Therefore, the
evolution of $p_i^t$ follows
\begin{equation}\label{eq1}
p_i^{t+1}=(1-p_i^t-p_i^{t-1})(I_i^t+(1-I_i^t)[1-\prod_{j=1}^{N_s}(1-p_j^tsA_{ij})]),
\end{equation}
where $I_i^t$ is the probability of sensor $i$ being activated by
infection at time $t$, $s$ is the coupling strength of sensors, and
$A$ is the adjacency matrix of the sensor network: $A_{ij}=1$ if
sensor $i$ and $j$ are connected, and $A_{ij}=0$ otherwise. The term
$(1-p_i^t-p_i^{t-1})$ is the probability that sensor $i$ is at
resting state at time $t$, while the term
$(1-I_i^t)[1-\prod_{j=1}^{N_s}(1-p_j^tsA_{ij})]$ is the probability
of sensor $i$ being activated by its neighboring sensors at time
$t$, rather than by infection.

To solve Eq.\ref{eq1}, we need to know the infection intensity
$I_i^t$. However, $I_i^t$ is highly related to sensor $i$'s
topological property. For instance, hubs are more likely to be
infected \cite{Albert2000,Pastor2001}. To eliminate the influence of
topological structure, we first adopt ER random networks in
simulations, and select sensors randomly. Under this condition, we
suppose the infection intensity $I_i^t$ for each sensor $i$ is
approximately the same. Because the sensor network is homogeneous,
we assume that the active probability $p_i^t$ is approximately the
same for all the sensors, i.e., $p_1^t\approx\cdots\approx
p_{N_s}^t\approx F^t$. In addition, using mean-field approximation,
the term $\prod_{j=1}^{N_s}(1-p_j^tsA_{ij})$ can be estimated by
$(1-F^ts)^{\langle k\rangle}$. Consequently, Eq.\ref{eq1} is
simplified to
\begin{equation}\label{eq2}
F^{t+1}=(1-F^t-F^{t-1})(I^t+(1-I^t)[1-(1-F^ts)^{\langle k\rangle}]).
\end{equation}
Because the sensor network is at the critical state, we have
$\langle k\rangle s=1$. For small-scale spreading, $F^t$ is close to
zero. To second order, the term $[1-(1-F^ts)^{\langle k\rangle}]$
can be approximated by $F^t+(\langle k\rangle-1)(F^t)^2/2\langle
k\rangle$. Then, Eq.\ref{eq2} is further reduced to
\begin{equation}\label{eq3}
F^{t+1}=I^t+(1-2I^t)F^t-I^tF^{t-1}-C(1-I^t)(F^t)^2-(1-I^t)F^tF^{t-1},
\end{equation}
where $C=(3\langle k\rangle-1)/2\langle k\rangle$.

Equation \ref{eq3} is an iterative updating function of $F^t$. In
fact, $F^{t+1}$ depends on $F^t$, $F^{t-1}$ and $I^t$. The infection
intensity $I^t$ is governed by the spreading dynamics. In
homogeneous networks, the evolution of $I^t$ satisfies specific
differential equations for SIR, SIS, Rumor and SIRL models, as shown
in the Materials and Methods section. By solving these equations, we
can calculate the infection intensity $I^t$. For SIR, SIS and SIRL
spreading, $I^t$ is the proportion of infected people $i(t)$. For
Rumor model, $I^t$ is the density of spreaders $s(t)$. With this
information, we calculate the response of the sensor network by
using the initial conditions $F^0=0$, $F^1=I^0$. In the cases of
SIR, Rumor and SIRL models, iterations continue until $I^t$ becomes
zero. For SIS model, iteration stops after $I^t$ becomes steady. The
influence of spreading can also be obtained by these equations. For
SIR, Rumor and SIRL models, the influence $M$ is the value of $r(t)$
when there are no infected people or spreaders. The influence $\rho$
for SIS model is the value of $i(t)$ in the steady state.

\begin{figure}%[!ht]
\begin{center}
\includegraphics[width=4in]{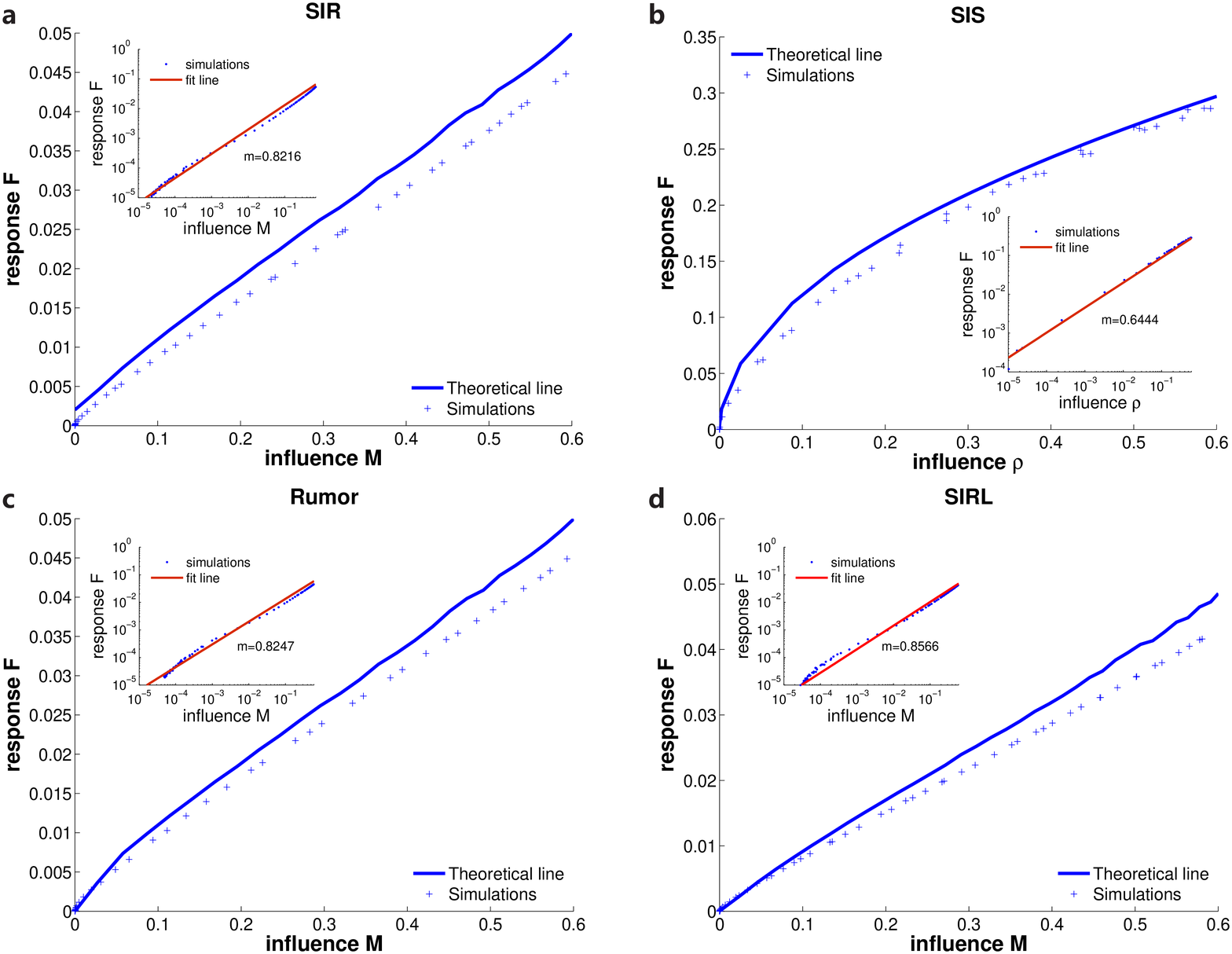}
\end{center}
\caption{ {\bf Theoretical analysis of the dynamics of excitable
sensors.} For SIR, SIS, Rumor and SIRL models, we display the
relationship between the response and influence in (a), (b), (c) and
(d) respectively. We adopt ER random networks with size $10^5$ and
average degree $10$. $10\%$ of nodes are randomly selected to be
sensors, which are connected in a homogeneous random network with
average degree $\langle k\rangle=10$. Solid lines are theoretical
predictions and cross symbols represent simulation values. In
simulations, we vary infection rate $\beta$ and keep $\mu=1$ for all
models. The contacting ability $L$ in SIRL model is set to be 5.
Insets show the power-law fit of the data, where $m$ is the
power-law exponent. } \label{fig2}
\end{figure}

To verify our theoretical analysis, we run simulations on ER random
networks. We generate ER random networks with size $10^5$ and
average degree $10$. Then, we randomly choose $10\%$ of nodes as
sensors, forming a random network with average degree $\langle
k\rangle=10$. In each simulation, we randomly select a node as the
spreading source. For each spreading model, we calculate the
theoretical values of response and influence for different infection
rates $\beta$ as we have explained above. In Fig.\ref{fig2}, the
theoretical lines agree well with the simulation results for all
considered spreading dynamics. Moreover, we find that the response
$F$ follows a power-law relation with influence, which is displayed
in the insets of Fig.\ref{fig2}. Clearly, the power-law exponent $m$
depends on spreading dynamics.

In terms of heterogeneous networks, we need to calculate $I_i^t$ for
each sensor $i$. Given the adjacency matrix $\bar{A}$ of the social
network, one can obtain $I_i^t$ for each spreading model by
iterating corresponding evolution equations, which are given in
Supporting Information S1. Combining the information of $I_i^t$ and
Eq.\ref{eq1}, we are able to calculate the theoretical values of
response and influence. In simulations, we generate BA scale-free
networks with size $10^5$ and average degree $10$
\cite{Barabasi1999}. The simulation results and theoretical values
are presented in S1 Fig.S1. The analytical analysis can well predict
simulation results.

\subsection*{Performance on spreading models }

There are several commonly used strategies to place sensors in
social networks. We introduce four of them to test the efficacy of
excitable sensor networks. The most straightforward method is to
randomly select nodes as detecting sensors. Another heuristic
alternative is to select hubs as sensors. It has been shown that
hubs are easily infected during an outbreak because they possess
large numbers of connections to other nodes
\cite{Albert2000,Pastor2001}. Therefore, we can pick nodes with
degree ranking in top $f$ percent as sensors to catch spreading
instances of small influence. The third method is to monitor the
friends of randomly selected individuals \cite{Christakis2010}. It
is known that individuals located in the center of networks are more
likely to be infected \cite{Kitsak2010,Pei2014}. Generally speaking,
the neighbors of a randomly chosen person tend to have larger number
of connections and higher $k$-shell indices \cite{Christakis2010}.
Moreover, this strategy is applicable even when we lack complete
information about the topological structures. Apart from these
sensor placing strategies, another important approach is based on
the distance centrality or the Jordan center of a graph
\cite{Lokhov2014,Comin2011,Shah2010,Shah2011}. For a graph $G$, the
distance centrality of node $i\in G$, $D(i, G)$, is defined as $D(i,
G)=\sum_{j\in G}d(i, j)$, where $d(i, j)$ is the shortest path
distance from node $i$ to node $j$ \cite{Shah2011}. Intuitively,
nodes with smaller distance centrality are closer to other nodes.
Consequently, nodes with small distance centrality are selected as
sensors in this strategy. After placing the sensors, we monitor the
statuses of these sensors during a spreading process. The influence
of spreading is detected as the proportion of infected sensors in
the cases of SIR, Rumor and SIRL models or the fraction of sustained
infection in the steady state for SIS model. In the following
analysis, we refer to these four strategies as {\it random}, {\it
targeted}, {\it acquaintance} and {\it distance} methods,
respectively.

To compare the efficacy of excitable sensor networks with other
methods, we apply spreading dynamics on three different types of
real-world social networks: an online social network - facebook
\cite{Viswanath2009}, a coauthorship network of scientific
publications \cite{Yang2012}, and a communication network of emails
\cite{Leskovec2009}. Explanations and details of the networks can be
found in the Materials and Methods section. The selected networks
are representative in their corresponding domains, and are widely
used in previous studies of social networks. Therefore, these
networks can reflect the characteristics of social networks in real
life. In simulations, because nodes with more connections have a
larger chance to be infected \cite{Albert2000,Pastor2001}, we select
nodes ranking in top $f$ percent in degree as sensors in the
excitable sensor network.

\begin{figure}%[!ht]
\begin{center}
\includegraphics[width=4in]{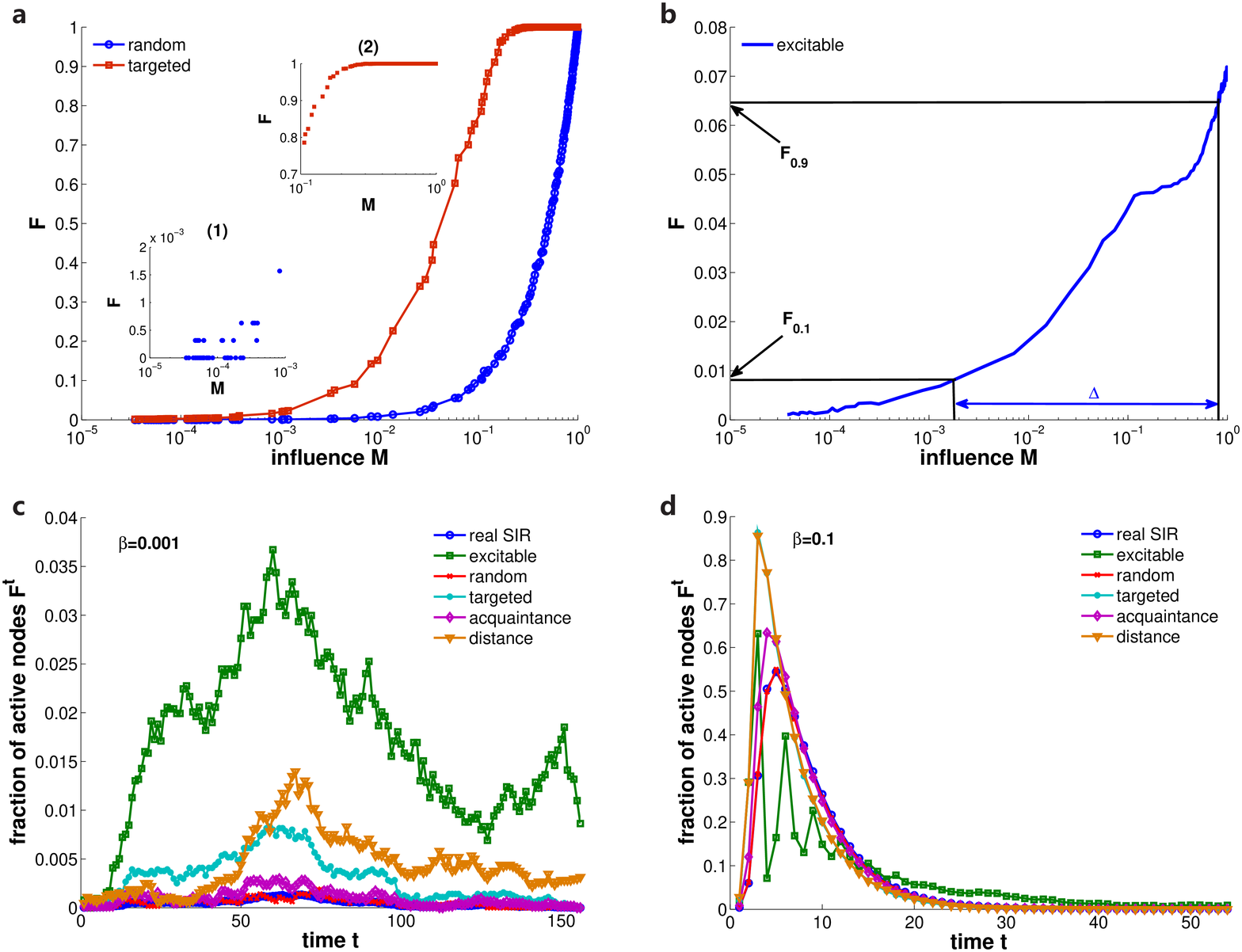}
\end{center}
\caption{ {\bf Response of sensor networks to SIR epidemic
spreading.} Here we run SIR model on facebook social network and
select $10\%$ of nodes as sensors. The average degree of sensor
network is set as $\langle k\rangle=4$, so the coupling strength of
excitable sensors is $s=0.25$. We set $\mu=0.2$ in simulations. The
source is selected as a hub with degree $k=1089$. (a) The random
sensors fail to detect small-scale epidemic spreading and targeted
sensors saturate only after the spreading occupies about $20\%$ of
population. (b) The excitable sensor network is capable of detecting
small-scale epidemic spreading and distinguishing large-scale
spreading. Straight lines indicate relevant parameters to calculate
the dynamic range $\Delta$. (c) and (d) display the fraction of
active sensors $F^t$ for different methods when we set $\beta=0.001$
and $\beta=0.1$ in SIR modeling respectively. } \label{fig3}
\end{figure}

Fig.\ref{fig3}(a) displays the responses of random and targeted
sensors to SIR epidemic spreading. By varying the infection
probability $\beta$, we can create diffusion instances with various
influence. We select $10\%$ of nodes in facebook social network as
sensors. The response of random sensors follows a linear
relationship with influence. In the range of small influence, random
sensors fail to detect some small-scale spreading and, more
importantly, cannot distinguish the influence clearly. This
phenomenon is better illustrated in the inset of Fig.\ref{fig3}(a).
Many spreading processes have response $F=0$, and large numbers of
diffusion processes with distinct influence produce same response.
In contrast to random strategy, targeted sensors perform better for
small influence. The response is amplified because of the
topological property of sensors, and the influence range that can
trigger distinguishable responses is extended to the left by at
least one order of magnitude. However, in the range of large
influence, the response of targeted sensors saturates only when
approximately $20\%$ of individuals are infected. For spreading
processes with influence larger than $20\%$, the response of
targeted sensors is always $1$. This dramatically diminishes the
effective detecting range in which we can rank the influence
correctly.

We now examine the performance of excitable sensor networks. In
Fig.\ref{fig3}(b), the response curve of the excitable sensor
network is displayed. The sensors are capable of detecting
small-scale epidemic spreading and distinguishing large-scale
diffusion. This can be explained by Fig.\ref{fig3}(c) and (d). In
the case of small influence, we set the infection probability
$\beta=0.001$. Compared with other strategies, the fraction of
active sensors $F^t$ for the excitable sensor network is greatly
amplified because of the signal transmission among sensors. The
random strategy reflects the real proportion of infected people at
each time step. Targeted, distance and acquaintance strategies
exploit sensors' relatively central locations in the social network,
thus enhancing the infected population of sensors. The excitable
sensor network relies on both sensors' topological advantages and
signal transmission processes. In this way, small-scale spreading
instances are more likely to be detected by the excitable sensor
network. When the response is amplified, it makes the influence more
distinguishable. For large-scale diffusion shown in
Fig.\ref{fig3}(d), where we set $\beta=0.1$, a peak of infection
appears. The fraction of active sensors for targeted strategy
becomes the largest, which produces the early saturation of targeted
sensors. In contrast, oscillations emerge in the excitable sensor
network because of the self-inhibition effect. Once a large fraction
of sensors are activated, they update into refractory state.
Therefore, in the next time step, $F^t$ will drop dramatically.
After these sensors changing back to resting state, they can be
activated again and create another crest. This phenomenon prevents
the early saturation of sensor networks. As a consequence, the
excitable sensor network can correctly detect the influence of
large-scale spreading.

\begin{figure}%[!ht]
\begin{center}
\includegraphics[width=4in]{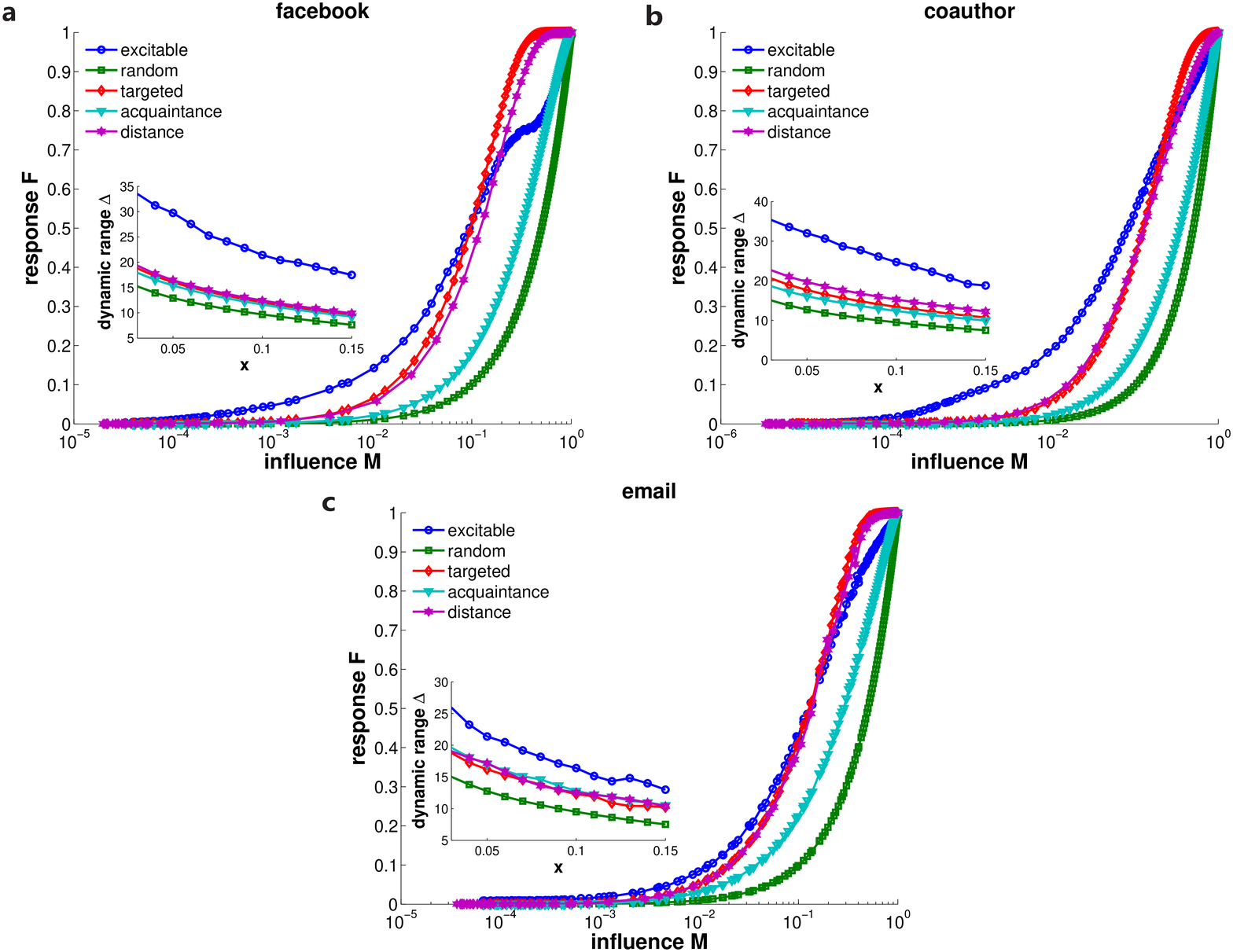}
\end{center}
\caption{ {\bf Comparison of performances of different strategies in
response to SIR spreading dynamics.} We apply SIR model on facebook
(a), coauthor (b) and email (c) social networks, and display the
response curve for each strategy. $10\%$ of nodes are selected as
sensors. We construct an excitable sensor network with average
degree $\langle k\rangle=4$, and set $\mu=0.2$ in simulations. The
sources are selected as hubs with degree $k=1089$, $343$ and $1383$
respectively. The response curves for all cases are normalized to
the unit interval $[0, 1]$. The insets show the dynamic range for
each case when we vary the calculation interval $[F_x,F_{1-x}]$ from
$x=0.01$ to $x=0.15$. } \label{fig4}
\end{figure}

To evaluate the results for other types of social networks, we also
apply SIR model on coauthor and email social networks.
Fig.\ref{fig4} displays the comparison of the performances of
different strategies in response to SIR spreading dynamics for
facebook, coauthor and email networks. In simulations, we select
$10\%$ of nodes as sensors and choose hubs as diffusion sources. We
construct an excitable sensor network with average degree $\langle
k\rangle=4$, and set $\mu=0.2$. To compare the response curves
directly, we normalize the response for each strategy to the unit
interval $[0,1]$. For all considered social networks, the response
curves share common properties: in the range of small influence, the
response of excitable sensors stays larger than those of other
methods, while it is suppressed for large-scale diffusion. This
leads to a broader effective detection range in which spreading
processes can be detected and ranked reliably. We calculate the
dynamic range for each case by varying the interval $[F_x,F_{1-x}]$
from $x=0.01$ to $x=0.15$. The insets show that the dynamic range of
the excitable sensor network is considerably higher than those of
random, targeted, acquaintance and distance strategies. This fact
directly indicates the better performance of excitable sensor
networks in detecting the influence of SIR spreading. In addition,
random sensors perform worst because they are incapable of
amplifying small-scale spreading.

Although we have tested the efficacy of the excitable sensor network
for SIR spreading dynamics, it is still desirable to evaluate its
performance for other spreading mechanisms. In Supporting
Information SIS, we perform SIS, Rumor and SIRL dynamics on
facebook, coauthor and email social networks, obtaining similar
results in S1 Figs.S2-S4. All evidence supports that the excitable
sensor network outperforms random, targeted, acquaintance and
distance strategies. A general conclusion we obtain from the
analysis is that the dynamic range of the excitable sensor network
is higher for all considered spreading dynamics (SIR, SIS, Rumor and
SIRL) and social networks (facebook, coauthor and email networks).
The consistency of the results stems from the inherent property of
excitable media. According to the dynamics of excitable elements,
excitable networks can amplify weak stimuli and suppress intense one
as well. This functional characteristic is independent of which
spreading dynamics or social networks we adopt. Consequently, it is
potentially applicable to various spreading dynamics and topological
structures.

\begin{figure}%[!ht]
\begin{center}
\includegraphics[width=4in]{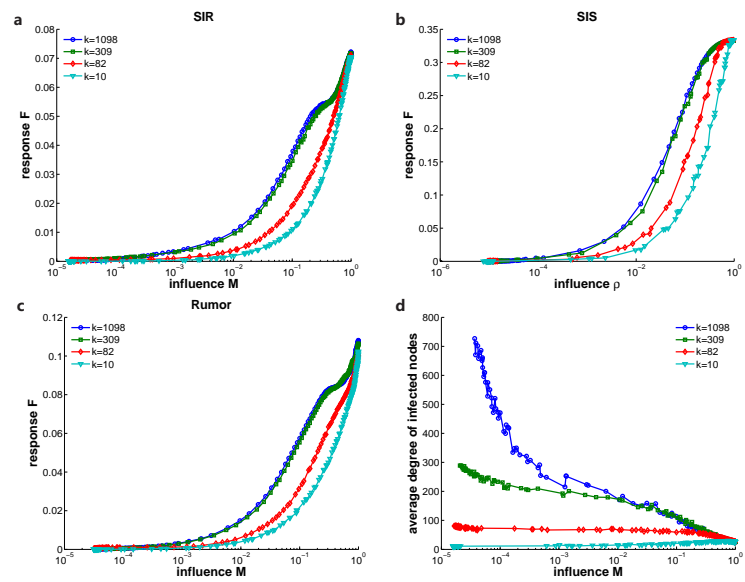}
\end{center}
\caption{ {\bf Response curves of excitable sensor networks for
different spreading sources.} The SIR (a), SIS (b) and Rumor (c)
spreading models are applied on facebook social network. Four
distinct nodes are selected as diffusion sources. The selected
sources have degree $k=1089$, $309$, $82$ and $10$. The relationship
between the response and spreading influence is presented. In (d) we
plot the average degree of infected people versus the spreading
influence for SIR model originating from different sources. }
\label{fig5}
\end{figure}

After we have verified the effectiveness of excitable sensor
networks for various spreading models and social networks, we
evaluate how excitable sensors respond to spreading processes
originating from different sources. We select four distinct nodes in
facebook social network as spreading origins. The selected nodes
have degrees $k=1089$, $309$, $82$ and $10$, representing distinct
groups of users. We run SIR, SIS and Rumor models on facebook social
network originating from these sources. The relationship between the
responses of excitable sensor networks and spreading influence is
presented in Fig.\ref{fig5}. Obviously, the response curves for
sources with smaller degrees stay lower. The amplifying effect of
excitable sensors is more effective for hubs in the region of small
influence and decreases as the degree of source diminishes. The
disparity in the amplifying effect can be explained by the property
of infected nodes in spreading dynamics. Despite the fact that the
spreading processes have same number of infected people, the
property of these individuals varies significantly for different
sources. In Fig.\ref{fig5}(d), we plot the average degree of
infected individuals versus spreading influence for SIR model
originating from four selected spreading sources. Clearly, in the
region of small influence, the average degree of infected people for
a well-connected source is far higher than that for a low-degree
source. As the spreading influence increases, this difference is
narrowed because more nodes are infected. Fig.\ref{fig5}(d)
indicates that for a spreading process originating from a
high-degree source, the infected people are also inclined to possess
large numbers of connections. This will cause more sensors to be
activated because we pick highly connected nodes as sensors. In this
way, the response for a high-degree source will become higher.
Because of this effect, excitable sensor networks are more likely to
detect spreading instances from high-degree sources under same
circumstances.

\subsection*{Effect of the construction method of excitable sensor networks}

After examining the performance of excitable sensor networks through
simulations, we would like to discuss the impact of the construction
method of sensor networks. First, we explain why the sensor network
is constructed as a homogeneous random network. It has been found
that homogeneous networks enhance the dynamic range more than
heterogeneous networks \cite{Kinouchi2006,Larremore2011}. In
ref\cite{Larremore2011}, the dynamic range at critical state
$\Delta$ can be predicted by
\begin{equation}\label{TheoDR}
\Delta=10\log_{10}\frac{2}{3F^2_*}-10\log_{10}\frac{\langle
vu^2\rangle}{\langle v\rangle\langle u\rangle^2},
\end{equation}
where $F_*$ is the lower threshold response, $u$ and $v$ are the
right and left dominant eigenvectors of sensor network's adjacency
matrix. Here $\langle vu^2\rangle=\sum_{i=1}^{i=N_s}v_iu_i^2/N_s$,
$\langle v\rangle=\sum_{i=1}^{i=N_s}v_i/N_s$, and $\langle
u\rangle=\sum_{i=1}^{i=N_s}u_i/N_s$. Given a fixed lower threshold
response $F_*$, only the term $-10\log_{10}(\langle
vu^2\rangle/(\langle v\rangle\langle u\rangle^2))$ can affect the
dynamic range $\Delta$. In our case of undirected networks where
$u_i=v_i$, the second term becomes $-10\log_{10}(\langle
u^3\rangle/\langle u\rangle^3)$. Since the entries of the dominant
eigenvector are first-order approximations to the degrees of
corresponding nodes ($u_i\approx k_i$) \cite{Larremore2011}, the
second term suggests that $\Delta$ should increase as the degree
distribution becomes more homogeneous.

In addition to theoretical analysis, we also compare the
performances of ER random and BA scale-free sensor networks through
simulations in Supporting Information S1. In facebook social
network, we select $10\%$ of nodes as sensors and construct ER and
BA networks with same average degree $\langle k\rangle=10$. The
coupling strength $s$ is adjusted to achieve the critical state for
both cases. Results in S1 Fig.S5 indicate that for all considered
spreading dynamics, ER sensor networks have higher dynamic ranges
consistently, which justifies our choice of homogenous structure of
sensor networks.

Apart from the impact of network structure, we also need to check
the effect of number of sensors, or the fraction of sensors $f$. We
conduct a sensitivity analysis on the number of sensors.
Specifically, we run SIR, SIS, Rumor and SIRL models on facebook
social network for $f$ ranging from $0.01$ to $0.1$. For all
spreading dynamics, the shape of response curves does not
dramatically change along with the number of sensors (see S1
Fig.S6). Meanwhile, the dynamic ranges almost remain unchanged for
different fractions of sensors $f$. This indicates that the choice
of sensor numbers would not affect our result significantly.

\subsection*{Validation by real diffusion data}

While we have verified the effectiveness of excitable sensor
networks for diverse spreading models, it still remains unknown how
our method performs for real-world spreading instances. There is
evidence showing that the spreading processes in reality cannot be
fully characterized by theoretical models \cite{Goel2012}.
Additionally, the spreading dynamics in social networks are also
greatly affected by other human-related factors, such as homophily
\cite{Centola2011,Aral2009}, activity
\cite{Iribarren2009,Muchnik2013}, social reinforcement
\cite{Centola2010}, and social influence bias \cite{Muchnik2013a}.
Under these circumstances, many researchers turn to explore
empirical diffusion data in various platforms. Here, we investigate
some empirical spreading instances from Twitter
\cite{McCreadie2012}, an online social networking and microblogging
service that has gained worldwide popularity. We intend to validate
the efficacy of excitable sensors using real-life information
spreading in Twitter. Details about the Twitter data can be found in
the Materials and Methods section.

\begin{figure}%[!ht]
\begin{center}
\includegraphics[width=4in]{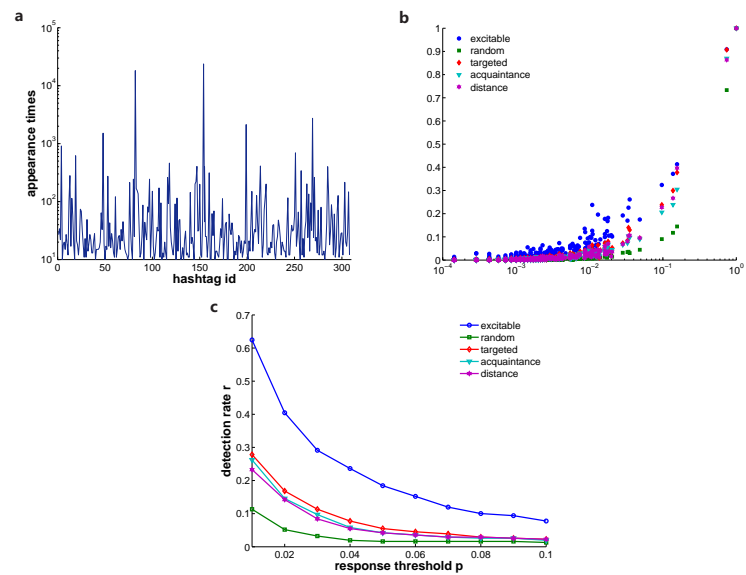}
\end{center}
\caption{ {\bf Validation of real diffusion instances in Twitter.}
The appearance frequency of selected 309 hashtags is shown in (a).
Hashtag ids are ranked chronologically. In (b) we present the
normalized response of different detecting strategies to these
hashtags. We also display the detection rate $r$ (the fraction of
detected hashtags) if we consider a topic is detected only when the
response is above a threshold $F_p$ in (c). } \label{fig6}
\end{figure}

To extract the information diffusion in Twitter, we examine the
contents of tweets. In Twitter, users usually include hashtags (a
word or an unspaced phrase prefixed with the number sign ``\#'') in
their tweets when they refer to specific topics. Therefore, tracking
the appearance of hashtags is a reliable method to infer the
information diffusion in Twitter. To be concrete, we examine the
tweets of 7014 users who have participated in discussions about the
protest in Egypt on January 25th, 2011. During a time window of 17
days, we filter 309 distinct hashtags appearing more than 10 times
among these users. We put a cutoff to omit small personal
discussions that rarely diffuse among users. The frequency of these
hashtags' appearing is displayed in Fig.\ref{fig6}(a), where hashtag
ids are ranked chronologically. The intensity of hashtags spans
several orders of magnitude. This property is suitable for testing
the performance of sensors in response to spreading instances that
vary by several orders of magnitude in influence.

To deploy sensors, we also require the structure of a social
network. In our case, we reconstruct the mention network of users
from the tweets to approximate the social network in real life. In
Twitter, mentions (``@username'') usually convey personal
conversations between users. Thus, we expect that the tie strength
of mentions is stronger than that of following relations
\cite{Grabowicz2012}. We do not track the follower network because
Twitter has imposed a strict limit on the access rate of Twitter
API, where we can obtain the list of followers for a given user.
Moreover, even though we can track the followers of these users, the
network structure may have changed significantly since the time of
data collection, and some users may have even closed their accounts
in Twitter. Considering all these limitations, we choose to use the
mention network, which can represent contemporary social relations
during the observation.

In detecting the topics in Twitter, we select $10\%$ of users as
sensors according to different strategies and form a sensor network
with average degree $\langle k\rangle=4$ for excitable sensors. We
regard each time step as a one-day interval. The state of a user at
each time step is determined by the content of his/her tweets. If a
user posts at least one tweet containing a specific hashtag during
one time step, we assume that this user is infected by this hashtag.
Otherwise, we assume that he/she is uninfected. During each time
step, in the excitable sensor network, sensors can be activated
either by infection, or by active neighboring sensors with the
probability of $s=0.25$ independently. The evolution dynamics remain
the same as the above simulations, but the spreading model in the
underlying social network is replaced by real diffusion instances.
Here, we note that how we construct the social network cannot affect
the spreading processes. All the information we use is ``at what
time, who has published a specific hashtag''. This information is
independent of the network topology. Therefore, our choice of social
network can only affect the selection of sensors and will not
strongly change our detection results.

Fig.\ref{fig6}(b) shows the responses of different detection
strategies. The influence of a hashtag is defined as the proportion
of users who have ever posted the hashtag during the observation
time. Each dot represents a response value for a specific hashtag.
Instead of following a smooth curve, the response dots are scattered
for all methods we consider. This phenomenon can be partially
explained by the distinction of spreading sources. Another factor
may be that multiple sources or independent spreaders exist in the
diffusion of hashtags \cite{Li2014}. In spite of this, the response
shows an increasing trend as the influence increases, which makes
large-scale diffusion distinguishable from small-scale ones. More
importantly, in the range of small influence, the response of the
excitable sensor network is higher on the whole. These results are
in accordance with our prediction through simulations.

To quantify the performance of each method, we define the detection
rate $r$ to measure how many spreading instances a method can
detect. Given a value $p\in[0,1]$, the detection rate $r(p)$ is
defined as the proportion of spreading instances whose responses are
larger than $F_p$, where $F_p=F_0+p(F_{max}-F_0)$. In other words,
if we assume that a topic is detected only when the response is
above $F_p$, the detection rate $r(p)$ quantifies the fraction of
topics that sensors can detect. A higher detection rate means a
better performance in detecting spreading. We present the detection
rate $r$ for $p\in[0.01,0.1]$ in Fig.\ref{fig6}(d). Obviously, the
detection rate of the excitable sensor network is the highest among
all the methods. This indicates that excitable sensor networks are
capable of detecting more spreading hashtags in real diffusion in
Twitter.

\section*{Conclusions}

The detection of spreading influence in social networks is an
important issue both in theory and practice. Inspired by the
mechanism by which the physical sensations of external stimuli
emerge, we propose a method to construct excitable sensor networks
in social networks. We study the dynamics of excitable sensors
analytically, and find the relationship between the response and
spreading influence. To compare the performance of our method with
other sensor placement strategies, we conduct SIR, SIS, Rumor and
SIRL simulations on three typical social networks. Because the
nonlinear amplification property of excitable elements is
independent of spreading dynamics, we obtain consistent conclusions:
under the same circumstances, excitable sensor networks exhibit
larger dynamic ranges, which implies that our method can react to
spreading processes in a wider range of influence. Excitable sensor
networks can not only detect small-scale spreading processes but
also better distinguish large-scale diffusions. In addition, if the
spreading processes originate from different sources, the average
degree of infected nodes is higher for high-degree origins. Because
of this fact, the excitable sensor network is more likely to detect
spreading instances originating from well-connected sources.
Moreover, we also validate the effectiveness of excitable sensor
networks with real diffusion data in Twitter. The detection result
shows that our method is capable of detecting more spreading topics.
Our approach provides a new route for designing effective detection
strategies.

% Do NOT remove this, even if you are not including acknowledgments
\section*{Acknowledgments}

%This work is supported by Major Program of National Natural Science Foundation of China (No. 11290141), NSFC (No. 11201018), International Cooperation Project
%No. 2010DFR00700, Fundamental Research of Civil Aircraft No. MJ-F-2012-04, and Innovation Foundation of BUAA for PhD Graduates.

%\section*{References}
% The bibtex filename
\bibliography{template}

\begin{thebibliography}{}

\bibitem{Anderson1992}
Anderson RM, May RM, Anderson B. Infectious diseases of humans:
dynamics and control. Oxford: Oxford University Press; 1992.

\bibitem{Diekmann2000}
Diekmann O, Heesterbeek JAP. Mathematical epidemiology of infectious
diseases. Chichester: Wiley; 2000.

\bibitem{Pastor-Satorras2014}
Pastor-Satorras R, Castellano C, Van Mieghem P, Vespignani, A.
Epidemic processes in complex networks; 2014. Preprint. Available:
arXiv:1408.2701.

\bibitem{Yan2014}
Yan S, Tang S, Pei S, Jiang S, Zheng Z. Dynamical Immunization
Strategy for Seasonal Epidemics. Phys Rev E 2014; 90: 022808.

\bibitem{Kumar2004}
Kumar R, Novak J, Raghavan P, Tomkins A. Structure and evolution of
blogspace. Communications of the ACM 2004; 47: 35-39.

\bibitem{Liben2008}
Liben-Nowell D, Kleinberg J. Tracing information flow on a global
scale using Internet chain-letter data. PNAS 2008; 105: 4633-4638.

\bibitem{Leskovec2009a}
Leskovec J, Backstrom L, Kleinberg J. Meme-tracking and the dynamics
of the news cycle. In Proceedings of the 15th ACM SIGKDD
international conference on Knowledge discovery and data mining.
2009; 497-506.

\bibitem{Yan2013}
Yan S, Tang S, Pei S, Jiang S, Zhang X, Ding W, et al. The spreading
of opposite opinions on online social networks with authoritative
nodes. Physica A 2013; 392: 3846-3855.

\bibitem{Watts2007}
Watts DJ, Peretti J, Frumin M. Viral marketing for the real world.
Harvard Business Rev 2007; May: 22-23.

\bibitem{Domingos2005}
Domingos P. Mining social networks for viral marketing. IEEE
Intelligent Systems 2005; 20: 80-82.

\bibitem{Leskovec2007}
Leskovec J, Adamic LA, Huberman BA. The dynamics of viral marketing.
ACM Transactions on the Web (TWEB) 2007; 1: 5.

\bibitem{Moreno2011}
Gonz\'{a}lez-Bail\'{o}n S, Borge-Holthoefer J, Rivero A, Moreno Y.
The dynamics of protest recruitment through an online network. Sci
Rep 2011; 1: 197.

\bibitem{Ginsberg2008}
Ginsberg J, Mohebbi MH, Patel RS, Brammer L, Smolinski MS, et al.
Detecting influenza epidemics using search engine query data. Nature
2008; 457: 1012-1014.

\bibitem{Culotta2010}
Culotta A. Towards detecting influenza epidemics by analyzing
Twitter messages. In Proceedings of the first workshop on social
media analytics 2010; 115-122.

\bibitem{Aramaki2011}
Aramaki E, Maskawa S, Morita M. Twitter catches the flu: detecting
influenza epidemics using Twitter. In Proceedings of the Conference
on Empirical Methods in Natural Language Processing 2011; 1568-1576.

\bibitem{Wilson2009}
Wilson K, Brownstein JS. Early detection of disease outbreaks using
the Internet. Canadian Medical Association Journal 2009; 180:
829-831.

\bibitem{Leskovec2007a}
Leskovec J, Krause A, Guestrin C, Faloutsos C, VanBriesen J, et al.
Cost-effective outbreak detection in networks. In Proceedings of the
13th ACM SIGKDD international conference on Knowledge discovery and
data mining 2007; 420-429.

\bibitem{Christakis2010}
Christakis NA, Fowler JH. Social network sensors for early detection
of contagious outbreaks. PLoS ONE 2010; 5: e12948.

\bibitem{Lokhov2014}
Lokhov AY, M\'{e}zard A, Ohta H, Zdeborov\'{a} L. Inferring the
origin of an epidemic with a dynamic message-passing algorithm. Phys
Rev E 2014; 90: 012801.

\bibitem{Pinto2012}
Pinto PC, Thiran P, Vetterli M. Locating the source of diffusion in
large-scale networks. Phys Rev Lett 2012; 109: 068702.

\bibitem{Chen2014}
Chen DB, Xiao R, Zeng A. Predicting the evolution of spreading on
complex networks. Sci Rep 2014; 4: 6108.

\bibitem{Keeling2004}
Keeling MJ, Brooks SP, Gilligan CA. Using conservation of pattern to
estimate spatial parameters from a single snapshot. PNAS 2004; 101:
9155-9160.

\bibitem{Brockmann2013}
Brockmann D, Helbing D. The hidden geometry of complex,
network-driven contagion phenomena. Science 2013; 342: 1337-1342.



\bibitem{Kinouchi2006}
Kinouchi O, Copelli M. Optimal dynamical range of excitable networks
at criticality. Nat Phys 2006; 2: 348-351.

\bibitem{Larremore2011}
Larremore DB, Shew WL, Restrepo JG. Predicting criticality and
dynamic range in complex networks: effects of topology. Phys Rev
Lett 2011; 106: 058101.

\bibitem{Hethcote2000}
Hethcote HW. The mathematics of infectious diseases. SIAM Rev 2000;
42: 599-653.

\bibitem{Pei2013}
Pei S, Makse HA. Spreading dynamics in complex networks. J Stat Mech
2013; 12: P12002.

\bibitem{Moreno2004}
Moreno Y, Nekovee M, Pacheco AF. Dynamics of rumor spreading in
complex networks. Phys Rev E 2004; 69: 066130.

\bibitem{Yang2007}
Yang R, Wang BH, Ren J, Bai WJ, Shi ZW, Wang WX, et al. Epidemic
spreading on heterogeneous networks with identical infectivity. Phys
Lett A 2007; 364: 189-193.

\bibitem{Yang2008}
Yang R, Zhou T, Xie YB, Lai YC, Wang BH. Optimal contact process on
complex networks. Phys Rev E 2008; 78: 066109.

\bibitem{Viswanath2009}
Viswanath B, Mislove A, Cha M, Gummadi KP. On the evolution of user
interaction in facebook. Proceedings of the 2nd ACM SIGCOMM Workshop
on Social Networks 2009; 37-42.

\bibitem{Yang2012}
Yang J, Leskovec J. Defining and evaluating network communities
based on ground-truth. Proceedings of the ACM SIGKDD Workshop on
Mining Data Semantics 2012; 3.

\bibitem{Leskovec2009}
Leskovec J, Lang KJ, Dasgupta A, Mahoney MW. Community structure in
large networks: Natural cluster sizes and the absence of large
well-defined clusters. Internet Mathematics 2009; 6: 29-123.


\bibitem{Barrat2008}
Barrat A, Barthelemy M, Vespignani A. Dynamical processes on complex
networks. Cambridge: Cambridge University Press; 2008.

\bibitem{Pastor2001}
Pastor-Satorras R, Vespignani A. Epidemic spreading in scale-free
networks. Phys Rev Lett 2001; 86: 3200.

\bibitem{Boguna2002}
Bogun\'{a} M, Pastor-Satorras R. Epidemic spreading in correlated
complex networks. Phys Rev E 2002; 66: 047104.

\bibitem{Kitsak2010}
Kitsak M, Gallos LK, Havlin S, Liljeros F, Muchnik L, Stanley HE, et
al. Identification of influential spreaders in complex networks. Nat
Phys 2010; 6: 888-893.

\bibitem{Kleinberg2007}
Kleinberg J. Cascading behavior in networks: Algorithmic and
economic issues. Algorithmic Game Theory 2007; 24: 613-632.

\bibitem{Lu2011}
L\"{u} L, Zhang YC, Yeung CH, Zhou T. Leaders in social networks,
the delicious case. PLoS ONE, 2011; 6: e21202.

\bibitem{Borge2012}
Borge-Holthoefer J, Moreno Y. Absence of influential spreaders in
rumor dynamics. Phys Rev E 2012; 85: 026116.

\bibitem{Larremore2011a}
Larremore DB, Shew WL, Ott E, Restrepo JG. Effects of network
topology, transmission delays, and refractoriness on the response of
coupled excitable systems to a stochastic stimulus. Chaos 2011; 21:
025117.

\bibitem{Pei2012}
Pei S, Tang S, Yan S, Jiang S, Zhang X, Zheng Z. How to enhance the
dynamic range of excitatory-inhibitory excitable networks. Phys Rev
E 2012; 86: 021909.

\bibitem{Larremore2014}
Larremore DB, Shew WL, Ott E, Sorrentino F, Restrepo JG. Inhibition
causes ceaseless dynamics in networks of excitable nodes. Phys Rev
Lett 2014; 112: 138103.

\bibitem{Albert2000}
Albert R, Jeong H, Barab\'{a}si AL. Error and attack tolerance of
complex networks. Nature 2000; 406: 378-382.

\bibitem{Barabasi1999}
Barab\'{a}si AL, Albert R. Emergence of scaling in random networks.
Science 1999; 286: 509-512.

\bibitem{Pei2014}
Pei S, Muchnik L, Andrade Jr JS, Zheng Z, Makse HA. Searching for
superspreaders of information in real-world social media. Sci Rep
2014; 4: 5547.

\bibitem{Comin2011}
Comin CH, da Fontoura Costa L. Identifying the starting point of a
spreading process in complex networks. Phys. Rev. E 2011; 84:
056105.

\bibitem{Shah2010}
Shah D, Zaman T. Detecting sources of computer viruses in networks:
theory and experiment. In ACM SIGMETRICS Performance Evaluation
Review 2010; 38: 203-214.

\bibitem{Shah2011}
Shah D, Zaman T. Rumors in a Network: Who's the Culprit? Information
Theory, IEEE Transactions on 2011; 57: 5163-5181.


\bibitem{Goel2012}
Goel S, Watts DJ, Goldstein DG. The structure of online diffusion
networks. In Proceedings of the 13th ACM conference on electronic
commerce 2012; 623-638.

\bibitem{Centola2011}
Centola D. An experimental study of homophily in the adoption of
health behavior. Science 2011; 334: 1269-1272.

\bibitem{Aral2009}
Aral S, Muchnik L, Sundararajan A. Distinguishing influence-based
contagion from homophily-driven diffusion in dynamic networks. PNAS
2009; 106: 21544-21549.

\bibitem{Iribarren2009}
Iribarren JL, Moro E. Impact of human activity patterns on the
dynamics of information diffusion. Phys Rev Lett 2009; 103: 038702.

\bibitem{Muchnik2013}
Muchnik L, Pei S, Parra LC, Reis SD, Andrade JS Jr, Havlin S,et al.
Origins of power-law degree distribution in the heterogeneity of
human activity in social networks. Sci Rep 2013; 3: 1783.

\bibitem{Centola2010}
Centola D. The spread of behavior in an online social network
experiment. Science 2010; 329: 1194-1197.

\bibitem{Muchnik2013a}
Muchnik L, Aral S, Taylor SJ. Social influence bias: A randomized
experiment. Science 2013; 341: 647-651.

\bibitem{McCreadie2012}
McCreadie R, Soboroff I, Lin J, Macdonald C, Ounis I, et al. On
building a reusable Twitter corpus. In Proceedings of the 35th
international ACM SIGIR conference on Research and development in
information retrieval 2012; 1113-1114.

\bibitem{Grabowicz2012}
Grabowicz PA, Ramasco JJ, Moro E, Pujol JM, Eguiluz VM. Social
features of online networks: The strength of intermediary ties in
online social media. PLoS ONE 2012; 7: e29358.

\bibitem{Li2014}
Li W, Tang S, Pei S, Yan S, Jiang S, Teng X, et al. The rumor
diffusion process with emerging independent spreaders in complex
networks. Physica A 2014; 397: 121-128.

\end{thebibliography}

%\\
%{\bf S1 Fig.S1 Theoretical analysis of the dynamics of excitable
%sensors in heterogeneous networks.} For SIR, SIS, Rumor and SIRL
%models, we display the theoretical values and simulation results of
%response.
%\\
%{\bf S1 Fig.S2 Response to SIS spreading dynamics of each method.}
%We run SIS model and display the normalized response curves for
%facebook, coauthor and email social networks.
%\\
%{\bf S1 Fig.S3 Performances of different strategies in response to
%Rumor dynamics.} The response curves for facebook, coauthor and
%email social networks are shown.
%\\
%{\bf S1 Fig.S4 Comparison of performances of different strategies in
%response to SIRL spreading dynamics.} We apply SIRL model on
%facebook, coauthor and email social networks, and display the
%relationship between the response and spreading influence.
%\\
%{\bf S1 Fig.S5 Impact of the topology of excitable sensor networks.}
%For facebook social network, we run SIR, SIS, Rumor and SIRL models
%with both ER random and BA scale-free sensor networks. The
%relationship between influence and response is displayed.
%\\
%{\bf S1 Fig.S6 Effect of the number of sensors.} We run SIR, SIS,
%Rumor and SIRL models on facebook social network for $f$ ranging
%from $0.01$ to $0.1$. The relationship between response and
%influence is displayed.

\section*{Figures}

\clearpage

\begin{flushleft}
{\Large \textbf{Supporting Information} }
\end{flushleft}

\section{Dynamics of excitable sensors in heterogeneous networks}

In the case of heterogeneous networks, we need to calculate $I_i^t$ for each sensor $i$. Since $I_i^t$ is significantly affected by sensor $i$'s topological feature, we have to use the individual-level evolution equations rather than the mean-field approximation. Assuming the adjacency matrix of the social network is $\bar{A}=\{\bar{A}_{ij}\}_{N\times N}$, we give the evolution differential equations for each spreading model as follows.

{\bf SIR model} For each node $i$ in the social network, denote the probability of node $i$ in susceptible, infected and recovered state at time $t$ as $S_i(t)$, $I_i(t)$ and $R_i(t)$ respectively. The state updating differential equations read
\begin{align*}
&  \frac{dI_i(t)}{dt}=[1-\prod_{j=1}^N(1-\beta\bar{A}_{ij}I_j(t)S_i(t))]-\mu I_i(t)\approx\beta S_i(t)\sum_{j=1}^N\bar{A}_{ij}I_j(t)-\mu I_i(t), \\
&  \frac{dS_i(t)}{dt}=-[1-\prod_{j=1}^N(1-\beta\bar{A}_{ij}I_j(t)S_i(t))]\approx-\beta S_i(t)\sum_{j=1}^N\bar{A}_{ij}I_j(t), \\
&   \frac{dR_i(t)}{dt}=\mu I_i(t).
\end{align*}

{\bf SIS model} Denote the probability of node $i$ in infected state at time $t$ as $I_i(t)$. The evolution of $I_i(t)$ follows
\begin{equation*}
  \frac{dI_i(t)}{dt}=[1-\prod_{j=1}^N(1-\beta\bar{A}_{ij}I_j(t)(1-I_i(t)))]-\mu I_i(t)\approx\beta (1-I_i(t))\sum_{j=1}^N\bar{A}_{ij}I_j(t)-\mu I_i(t).
\end{equation*}

{\bf Rumor model} Assume the probability of node $i$ in the state of spreader, ignorant and stifler at time $t$ as $S_i(t)$, $I_i(t)$ and $R_i(t)$ respectively. The evolution differential equations are
\begin{align*}
&  \frac{dI_i(t)}{dt}=-[1-\prod_{j=1}^N(1-\beta\bar{A}_{ij}S_j(t)I_i(t))]\approx-\beta I_i(t)\sum_{j=1}^N\bar{A}_{ij}S_j(t), \\
&  \frac{dS_i(t)}{dt}=[1-\prod_{j=1}^N(1-\beta\bar{A}_{ij}S_j(t)I_i(t))]-[1-\prod_{j=1}^N(1-\mu\bar{A}_{ij}(S_j(t)+R_j(t))S_i(t))] \\ & \approx\beta I_i(t)\sum_{j=1}^N\bar{A}_{ij}S_j(t)-\mu S_i(t)\sum_{j=1}^N\bar{A}_{ij}(S_j(t)+R_j(t)), \\
&   \frac{dR_i(t)}{dt}=[1-\prod_{j=1}^N(1-\mu\bar{A}_{ij}(S_j(t)+R_j(t))S_i(t))]\approx\mu S_i(t)\sum_{j=1}^N\bar{A}_{ij}(S_j(t)+R_j(t)).
\end{align*}

{\bf SIRL model} Denote the probability of node $i$ in susceptible, infected and recovered state at time $t$ as $S_i(t)$, $I_i(t)$ and $R_i(t)$. We have
\begin{align*}
&  \frac{dI_i(t)}{dt}=[1-\prod_{j=1}^N(1-\beta\bar{A}_{ij}\frac{L}{k_j}I_j(t)S_i(t))]-\mu I_i(t)\approx\beta LS_i(t)\sum_{j=1}^N\bar{A}_{ij}\frac{I_j(t)}{k_j}-\mu I_i(t), \\
&  \frac{dS_i(t)}{dt}=-[1-\prod_{j=1}^N(1-\beta\bar{A}_{ij}\frac{L}{k_j}I_j(t)S_i(t))]\approx-\beta LS_i(t)\sum_{j=1}^N\bar{A}_{ij}\frac{I_j(t)}{k_j}, \\
&   \frac{dR_i(t)}{dt}=\mu I_i(t).
\end{align*}
Here $k_j$ represents the degree of node $j$.

\begin{figure}%[!ht]
\begin{center}
\includegraphics[width=4in]{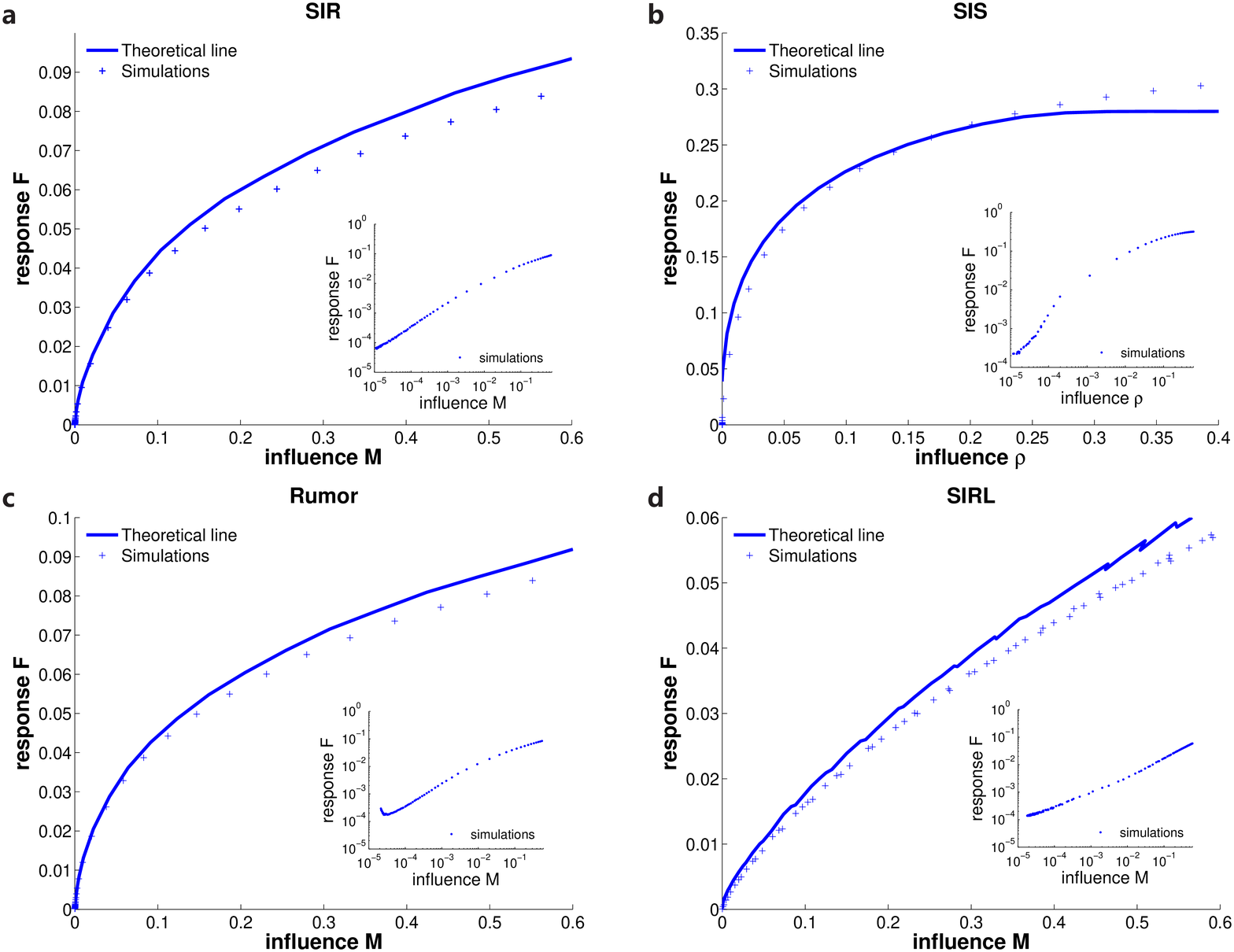}
\end{center}
\caption{
{\bf Theoretical analysis of the dynamics of excitable sensors in heterogeneous networks.} For SIR, SIS, Rumor and SIRL models, we display the relationship between the response and influence in (a), (b), (c) and (d) respectively. We adopt BA scale networks with size $10^5$ and average degree $10$. $10\%$ of nodes are randomly selected to be sensors, which are connected in a homogeneous random network with average degree $\langle k\rangle=10$. Solid lines are theoretical predictions and cross symbols represent simulation values. In simulations, we vary infection rate $\beta$ and keep $\mu=1$ for all models. The contacting ability $L$ in SIRL model is set to be 5.
}
\label{figS1}
\end{figure}

Combining the above equations and Eq.1 in the main text, we can calculate the theoretical values of the response and influence for different infection rates $\beta$. In order to validate the theoretical analysis, we perform simulations on heterogeneous networks. In particular, we generate BA scale-free networks with size $10^5$ and average degree $10$, and run each spreading model to obtain simulation results. The relationship between the theoretical lines and simulation results is displayed in Figure \ref{figS1} in S1 File. For all considered spreading dynamics, the theoretical lines agree well with the simulation results.

\section{Performance of excitable sensor networks for various spreading dynamics}

\begin{figure}%[!ht]
\begin{center}
\includegraphics[width=4in]{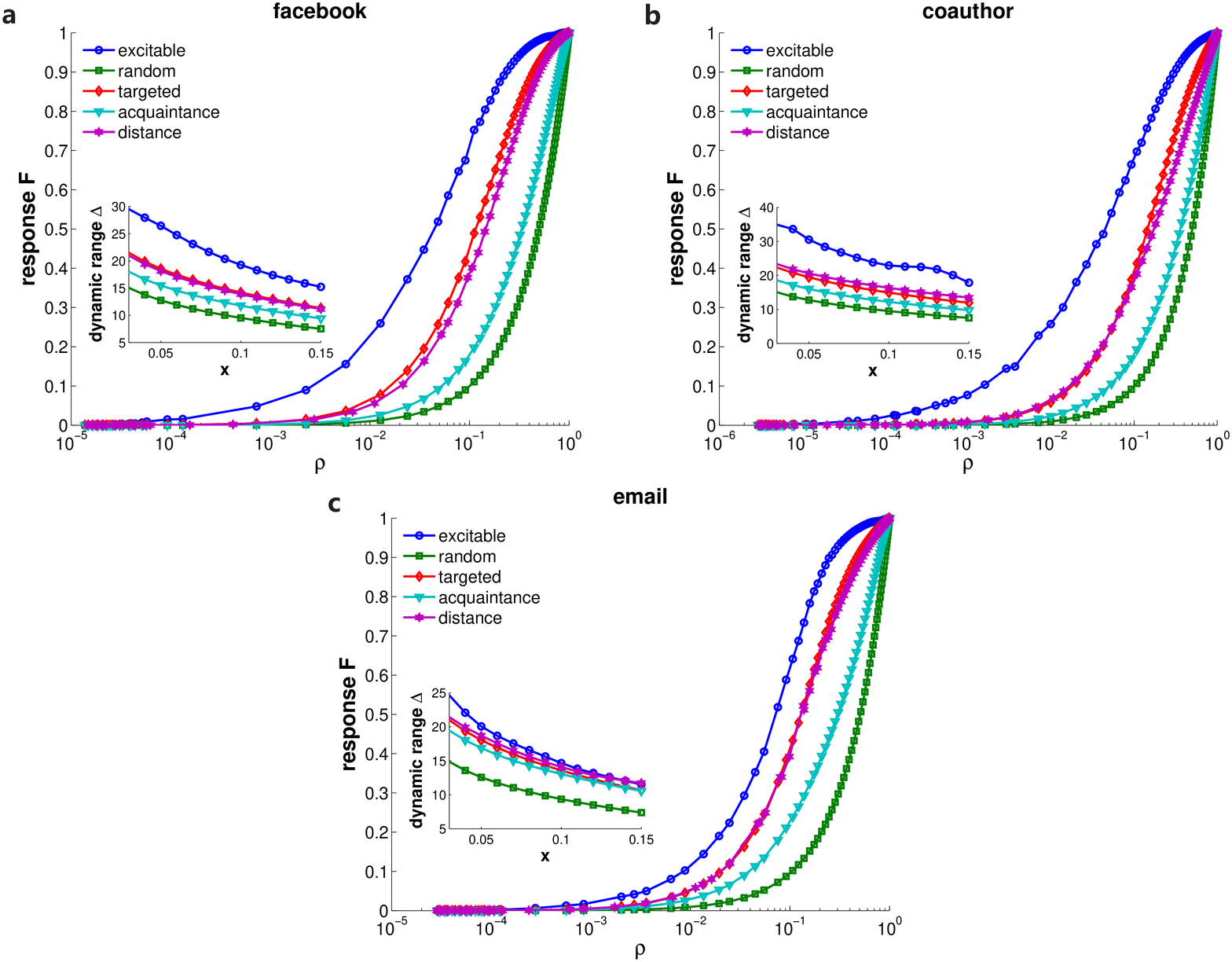}
\end{center}
\caption{
{\bf Response to SIS spreading dynamics of each method.} We run SIS model and display the normalized response curves for facebook (a), coauthor (b) and email (c) social networks. We select $10\%$ of nodes as sensors. The sources are selected as hubs with degree $k=1089$, $343$ and $1383$ respectively. The average degree of excitable sensor networks is $\langle k\rangle=4$, and $\mu=0.2$. The insets show the dynamic range for each case when we vary the calculation interval $[F_x,F_{1-x}]$ from $x=0.01$ to $x=0.15$.
}
\label{figS2}
\end{figure}

\begin{figure}%[!ht]
\begin{center}
\includegraphics[width=4in]{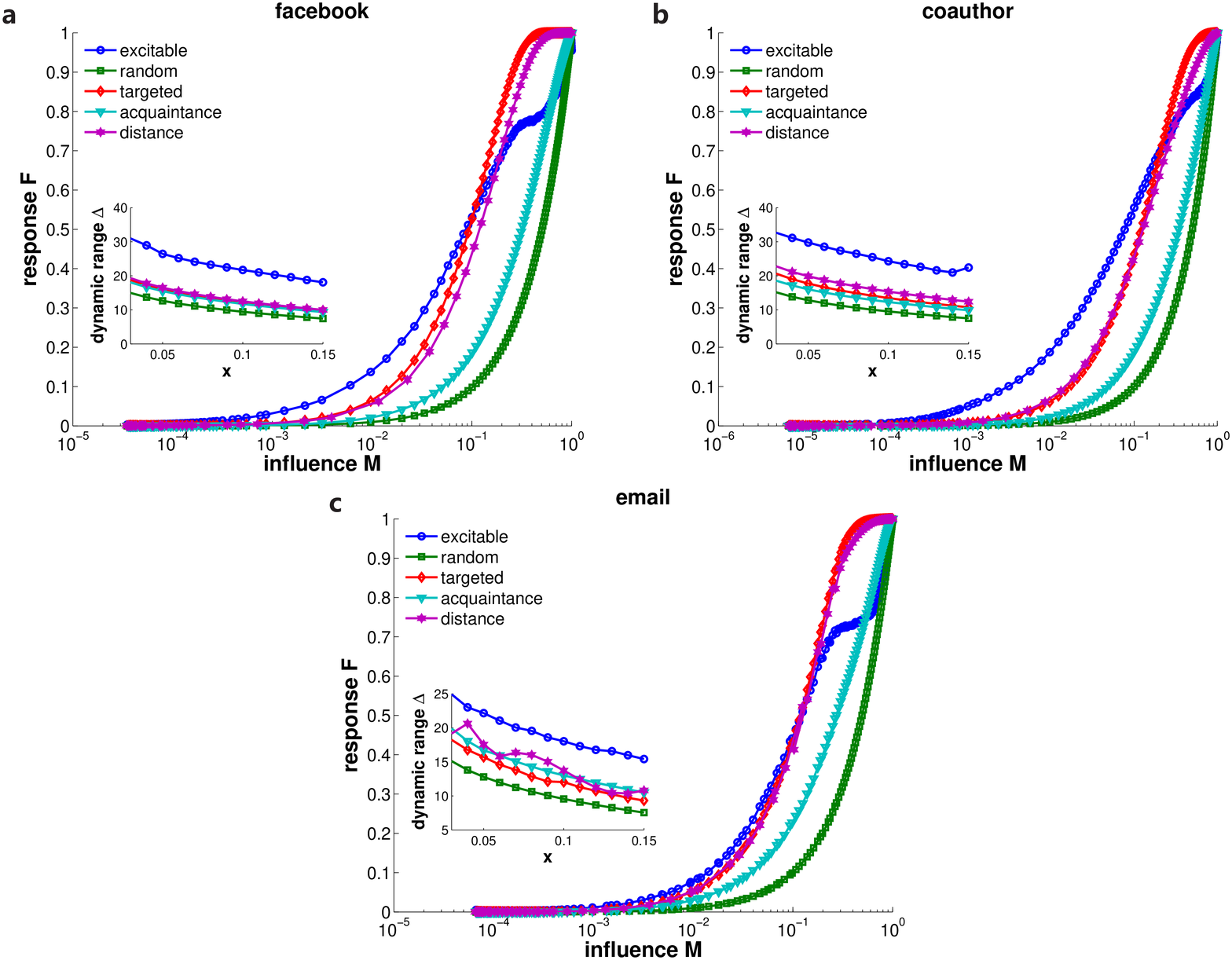}
\end{center}
\caption{
{\bf Performances of different strategies in response to Rumor dynamics.} The response curves for facebook (a), coauthor (b) and email (c) social networks are shown. We adopt Rumor dynamics and select source nodes with degree $k=1089$, $343$ and $1383$ respectively. Response curves are normalized to the unit interval $[0, 1]$. The excitable sensor networks contain $10\%$ of nodes, and have average degree $\langle k\rangle=4$. The parameter $\mu$ is set to be $1$. The dynamic range for each case is shown, where we vary the calculation interval $[F_x,F_{1-x}]$ from $x=0.01$ to $x=0.15$.
}
\label{figS3}
\end{figure}

\begin{figure}%[!ht]
\begin{center}
\includegraphics[width=4in]{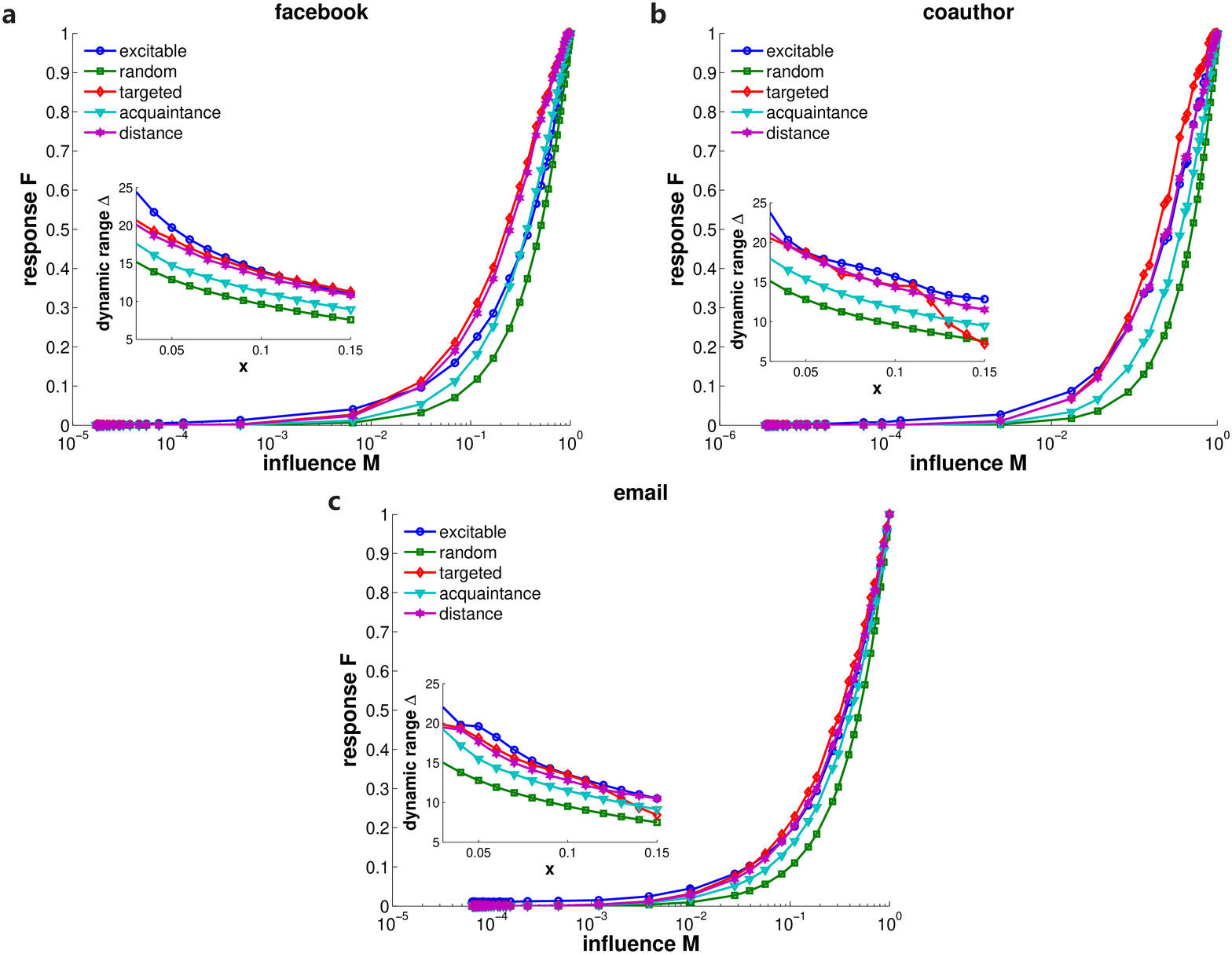}
\end{center}
\caption{
{\bf Comparison of performances of different strategies in response to SIRL spreading dynamics.} We apply SIRL model on facebook (a), coauthor (b) and email (c) social networks, and display the relationship between the response and spreading influence. $10\%$ of nodes are selected as sensors. We construct excitable sensor networks with average degree $\langle k\rangle=4$, and set $\mu=0.2$, $L=5$ in simulations. The sources are selected as hubs with degree $k=1089$, $343$ and $1383$ respectively. The response curves for all cases are normalized to the unit interval. The insets show the dynamic range for each case when we vary the calculation interval $[F_x,F_{1-x}]$ from $x=0.01$ to $x=0.15$.
}
\label{figS4}
\end{figure}

Although we have tested the efficacy of the excitable sensor network for SIR spreading dynamics in the main text, it is still desirable to evaluate its performance for other spreading mechanisms. To achieve this, we perform SIS, Rumor and SIRL dynamics on facebook, coauthor and email social networks, conducting similar analyses in Figures \ref{figS2}-\ref{figS4} in S1 File. Without loss of generality, we construct excitable sensor networks with average degree $\langle k\rangle=4$. We set $\mu=0.2$ for SIS and SIRL models and $\mu=1$ for Rumor model. The contacting ability $L$ in SIRL model is set to be 5. All simulations support that the excitable sensor network outperforms random, targeted, acquaintance and distance strategies.

\section{Effect of the construction method of excitable sensor networks}

\begin{figure}%[!ht]
\begin{center}
\includegraphics[width=4in]{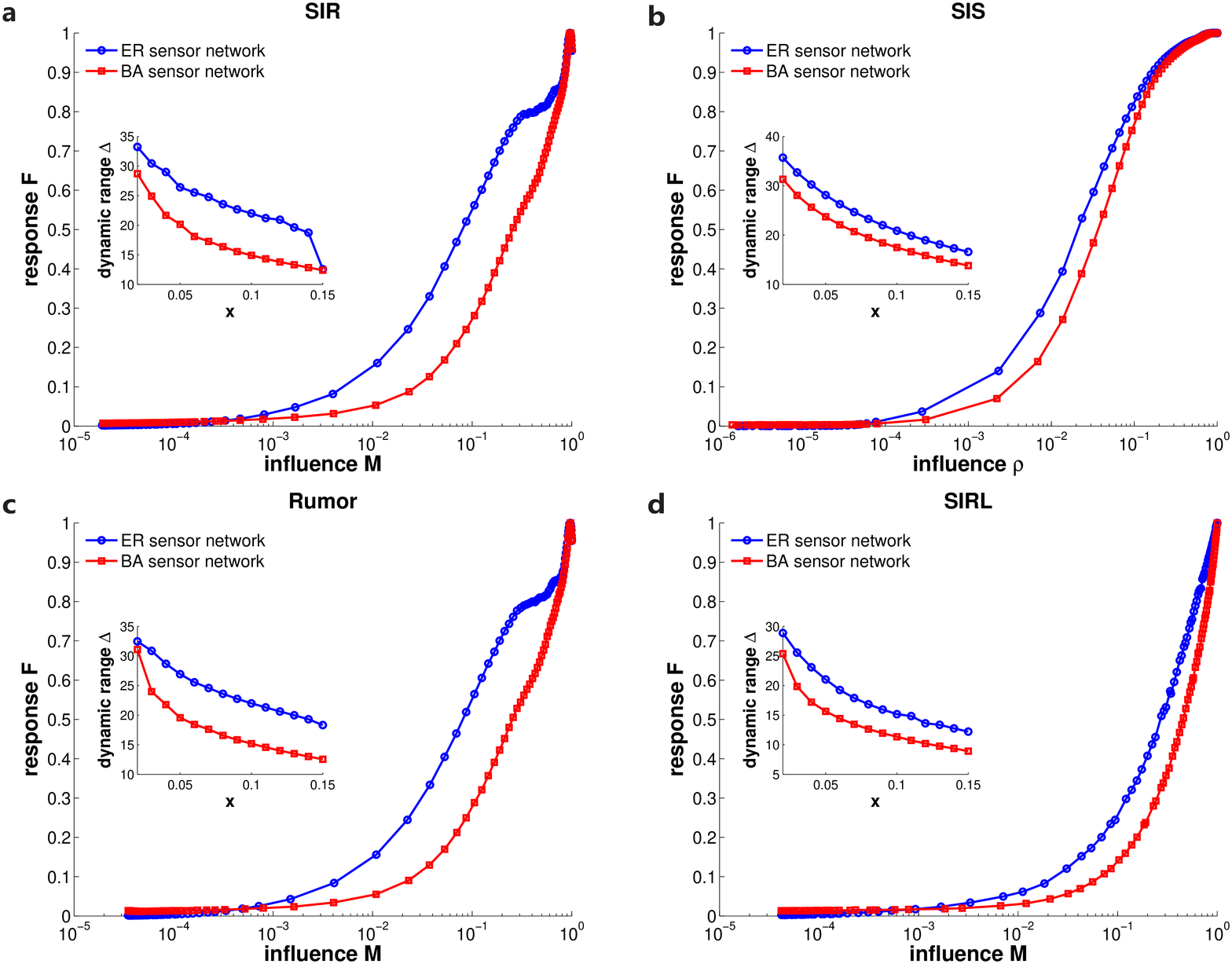}
\end{center}
\caption{
{\bf Impact of the topology of excitable sensor networks.} For facebook social network, we run SIR (a), SIS (b), Rumor (c) and SIRL (d) models with both ER random and BA scale-free sensor networks. The relationship between influence and response is displayed. The excitable sensor networks contain $10\%$ nodes and have average degree $\langle k\rangle=10$. We set $\mu=1$ for all models and $L=5$ for SIRL model. The source is selected as the hub with degree $k=1089$. The insets show the dynamic ranges when we vary the calculation interval $[F_x,F_{1-x}]$ from $x=0.01$ to $x=0.15$.
}
\label{figS5}
\end{figure}

To explore the impact of the topology of sensor networks, we run SIR, SIS, Rumor and SIRL models with both ER random and BA scale-free sensor networks. In facebook social network, we select $10\%$ nodes as sensors and construct ER and BA networks with the same average degree $\langle k\rangle=10$. The coupling strength $s$ is adjusted to achieve the critical state for both cases. For each BA scale-free sensor network, we first calculate the eigenvalue of the adjacency matrix $\lambda$, and then set the coupling strength $s=1/\lambda$. Results in Figure \ref{figS5} in SI File indicate that ER sensor networks have higher dynamic ranges.

\begin{figure}%[!ht]
\begin{center}
\includegraphics[width=4in]{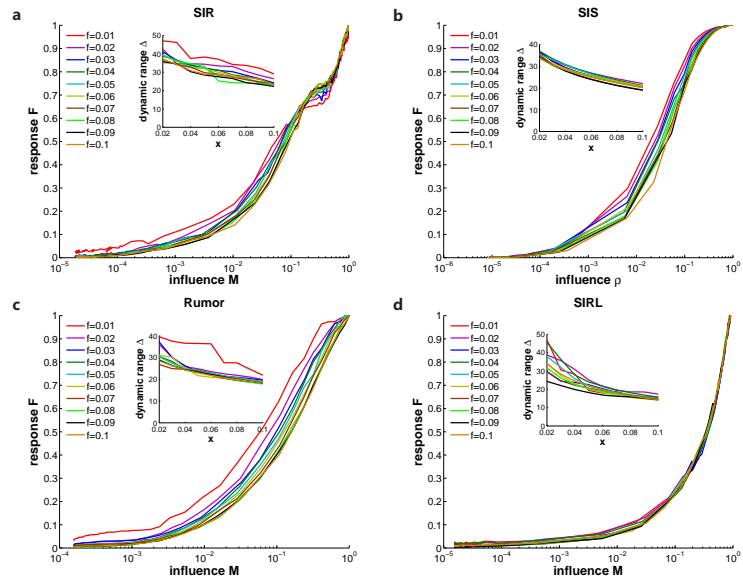}
\end{center}
\caption{
{\bf Effect of the number of sensors.} We run SIR (a), SIS (b), Rumor (c) and SIRL (d) models on facebook social network for $f$ ranging from $0.01$ to $0.1$. The relationship between response and influence is displayed. The excitable sensor networks have average degree $\langle k\rangle=4$. We set $\mu=1$ for all models and $L=5$ for SIRL model. The source is selected as the hub with degree $k=1089$. The insets show the dynamic range for each case when we vary the calculation interval $[F_x,F_{1-x}]$ from $x=0.01$ to $x=0.15$.
}
\label{figS6}
\end{figure}

In order to check the effect of $f$, we conduct a sensitivity analysis on the number of sensors. We simulate SIR, SIS, Rumor and SIRL models on facebook social networks for $f$ ranging from $0.01$ to $0.1$. In Figure \ref{figS6} in S1 File, the shape of response curves is not dramatically changed by the number of sensors. At the same time, the dynamic ranges almost remain unchanged for different fractions of sensors $f$. This indicates that the choice of sensor numbers would not affect our result significantly.

%\section*{Tables}
%\begin{table}[!ht]
%\caption{
%\bf{Table title}}
%\begin{tabular}{|c|c|c|}
%table information
%\end{tabular}
%\begin{flushleft}Table caption
%\end{flushleft}
%\label{tab:label}
% \end{table}

\end{document}